\documentclass{nature}
\usepackage{graphicx}
\makeatletter
\let\saved@includegraphics\includegraphics
\AtBeginDocument{\let\includegraphics\saved@includegraphics}
\renewenvironment*{figure}{\@float{figure}}{\end@float}
\makeatother

\usepackage{textcomp}
\usepackage{caption}

\title{Decoding electron tunnelling delay time by embracing wave-particle duality}

\author{Chuncheng Wang$^{1,6}$\footnote{These authors equally contribute  to this work.} $^{\dag}$, Xiaokai Li$^{1*}$, Xiwang Liu$^{2,3*}$, Jie Li$^{2}$, Shengpeng Zhou$^{1}$, Yizhang Yang$^{1}$, Xiaohong Song$^{2,3,4\dag}$, Jing Chen$^{5}$, Weifeng Yang$^{2,3,4\dag}$ \& Dajun Ding$^{1\dag}$}

\begin{document}
\captionsetup{ labelformat=empty}

\maketitle

\begin{affiliations}
 \item Institute of Atomic and Molecular Physics, and Advanced Light Field and Modern Medical Treatment Science and Technology Innovation Center of Jilin Province, Jilin University, Changchun 130012, China
 \item Research Center for Advanced Optics and Photoelectronics, Department of Physics, College of Science, Shantou University, Guangdong 515063, China
 \item Institute of Mathematics, Shantou University, Shantou, Guangdong 515063, China
 \item MOE Key Laboratory of Intelligent Manufacturing Technology, Shantou University, Shantou, Guangdong 515063, China
 \item Institute of Applied Physics and Computational Mathematics, P.O. Box 8009, Beijing 100088, China
 \item Laboratory for Physical Chemistry, ETH Z\"{u}rich, Vladimir-Prelog-Weg 2, 8093 Z\"{u}rich, Switzerland
\end{affiliations}

\begin{abstract}

Tunnelling lies at the heart of quantum mechanics and is a fundamental process in attosecond science, molecular biology, and quantum devices. Whether tunnelling takes time and how a microscopic particle transits through a barrier have been debated since the early days of quantum mechanics \cite{Maccoll1932,Hartman1962,1983Buttiker,Spielmann1994,Steinberg1995,Balcou1997,Landsman2014,Torlina2015,Camus2017}. The time required for an electron to tunnel through an atomic potential barrier has been measured with attosecond angular streaking (attoclock), and a recent work on the hydrogen atom claimed that electron tunnelling is instantaneous \cite{Sainadh2019}. However, the time required for Rb atoms to tunnel through an optical potential barrier has been measured to be on the order of milliseconds with a recent Larmor clock measurement \cite{2020nature}. The essence of electron and atom tunnelling is identical, but the reason for the contradictory conclusions remains unknown. Here, we demonstrate that the sub-barrier potential interaction is the root of the nonzero tunnelling delay time. We reveal that the wave-particle duality of a tunnelling electron must be fully taken into account when decoding the tunnelling delay time from the attoclock measurements. Based on energy-resolved attoclock measurements, we show that the tunnelling delay time of an electron ranges from 24 to 58 attoseconds, and the counterintuitive result that an electron with a lower energy may spend less time in the barrier is consistent with the velocity-dependent tunnelling time of atoms in Larmor clock measurements \cite{2020nature}. Our results unify the tunnelling time of microscopic particles by highlighting the classically forbidden sub-barrier potential interactions of matter waves.

\end{abstract}

Chronoscopy techniques, such as attoclock \cite{Eckle2008}, attosecond streak camera \cite{2010StreakingCamera}, RABBITT interferometry \cite{rabbitt}, attosecond transient absorption \cite{2010ATAS}, and high-harmonic spectroscopy \cite{2012HHG}, have allowed electron dynamics to be probed with attosecond resolution. Running counter to the recent progress of chronoscopy, the long-term debate on tunnelling time has recently intensified. To resolve the paradox between the atom and electron tunnelling, it is necessary to investigate the underlying physics of tunnelling and the methodology to measure the tunnelling time. In Larmor clock measurements for atom tunnelling, the clock ticks only when an atomic wave packet is in the optical potential barrier, and the nonzero tunnelling time is essentially the time of interaction between the atoms and the potential barrier \cite{2020nature}. Whereas, the tunnelling time of electrons has been measured using the attoclock, which employs a close-to-circularly-polarised femtosecond laser pulse to map time information into an angular offset of the most probable momentum vector with respect to the maximal vector potential. In contrast to the Larmor clock measurements, the Coulomb interaction between the electron and ionic core exists not only within the barrier but also in the continuum after the tunnelling, and both of these potential interactions might give rise to angular offsets in the attoclock measurements \cite{Eckle2008,Orlando2014,Torlina2015,Sainadh2019}. To extract the tunnelling time of electrons accurately, the contributions of Coulomb interactions during and after the tunnelling should be clearly distinguished, which is a formidable challenge for interpreting attoclock experiments.

To date, the widely adopted approaches for reconstructing the tunnelling time from attoclock measurements have relied on three assumptions \cite{Torlina2015}: (a) The highest probability for the electron tunnelling is at the peak of the electric field, which acts as a reference point of $t_{0}=0$. (b) The tunnelling would be completed at the moment, i.e. the ionization time $t_{exit}$, when an electron emerges in the continuum from the barrier. By causality, $t_{exit}$ is determined solely by the tunnelling process. If $t_{exit}>0$, there is a real tunnelling delay time $\tau=t_{exit}-t_{0}$ \cite{Hofmann2019}. By contrast, if $t_{exit}\approx0$, $\tau\approx0$ and instantaneous tunnelling would be implied. (c) After tunnelling, the electron dynamics in the continuum are described classically. These assumptions follow the sequentiality of events, treating the attoclock as a classical clock \cite{Pazourek2015}.

\begin{figure*}
	\begin{center}
		\includegraphics[width=\linewidth]{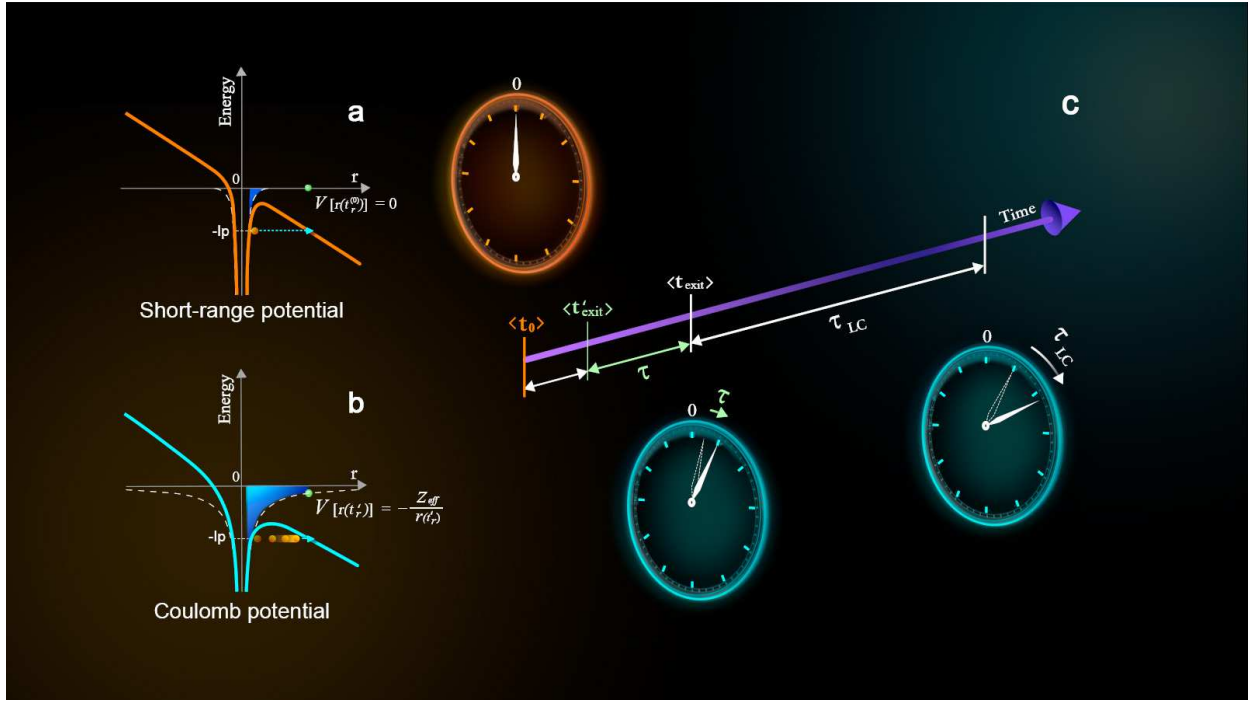}
	\end{center}
	\caption{\emph{\textbf{Fig.1$\mid$Electron tunnelling and tunnelling delay time.} The tunnelling of an electron
		is illustrated in the case of \textbf{a}, short-range potential and \textbf{b,} Coulomb potential. The white dashed lines in \textbf{a} and \textbf{b} are the field-free short-range and Coulomb potential, respectively. The blue shadows denote the nonzero potential interaction region during the tunnelling. \textbf{c,} The time components constitute the measured angular offset. $\tau_{LC}$ is the time induced by the Coulomb interaction in the continuum. See the main text for $\langle{t_0}\rangle$, $\langle{t^{\prime}_{exit}}\rangle$,  $\langle{t_{exit}}\rangle$ and $\tau$.}}

	\label{Figure1}
\end{figure*}

In a recent work on the hydrogen atom, to eliminate the Coulomb interaction in the continuum, a short-range potential instead of atomic potential was adopted in the numerical solution of the time-dependent Schr\"{o}dinger equation (TDSE), which leads to an angular offset of zero and the argument of instantaneous electron tunnelling \cite{Sainadh2019}. However, this substitution not only eliminated the Coulomb interaction in the continuum, but also diminished the sub-barrier potential interaction during the tunnelling, because the short-range potential rapidly approaches $0$ within the barrier region (see the white dashed curve in Figs. 1a and 1b). This would inevitably alter the sub-barrier potential interaction and the time a tunnelling particle spends inside the potential barrier, i.e. the tunnelling time.

Here, we perform energy-resolved electron attosecond angular streaking measurements (see S1 in Supplementary Information, SI) on Xe for comparison with velocity-dependent atom Larmor clock measurements \cite{2020nature}. The photoelectron momentum distribution (PMD) of tunnelling electrons driven by a 40 \emph{fs} laser pulse presents a series of above-threshold ionization (ATI) rings spaced by one-photon energy \cite{Agostini1979, Wang2019PRL}. The ATI rings exhibit different most probable angular offsets, which offer a unique way to explore the energy-dependence of electron tunnelling (Fig. 2a). The energy-resolved measurements are reproduced by the numerical solution of a three-dimensional TDSE. The simulated and experimental results reach quantitative agreement (Figs. 2d-h).

\begin{figure*}
	\begin{center}
		\includegraphics[width=\linewidth]{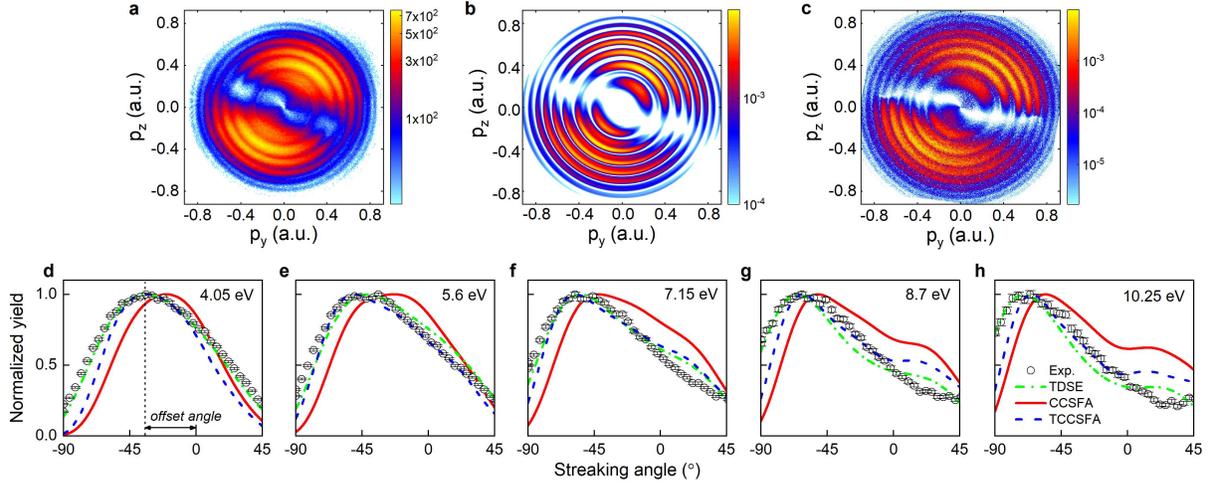}
	\end{center}
	\caption{\emph{\textbf{Fig.2$\mid$Energy-resolved attosecond angular streaking.} PMDs acquired from \textbf{a}, the measurements; \textbf{b}, the numerical solution of 3D-TDSE; and \textbf{c}, TCCSFA simulations.  Panels \textbf{d} through \textbf{h} show the measured and calculated angular distributions for ATI rings with different energies. The most probable angles for the experiment, TDSE and TCCSFA simulations reach quantitative agreement, which show clear angular offsets relative to the Coulomb-corrected strong-field approximation (CCSFA) calculations without the sub-barrier potential interaction. The peak laser intensity was (45$\pm$2) TW/cm$^{2}$ for the measurements and 45 TW/cm$^{2}$ for the simulations.}}
	\label{fig2}
\end{figure*}


To highlight the importance of sub-barrier potential interaction in the classically forbidden region, we first provide a consistent tunnelling-Coulomb-corrected strong-field approximation (TCCSFA, details in S2 of SI) method which fully includes the potential interactions both during and after tunnelling, and then compare the simulation results with that when the sub-barrier potential interaction is neglected. It has been thought previously that the sub-barrier potential interaction during the tunnelling hardly changes the final PMD \cite{Huismans2011}, however, here we clarify that it actually plays an essential role on the more accurate energy-resolved attoclock experiment: only when the sub-barrier potential interaction is taken into account, the simulations  can quantitatively reproduce the angular offsets in both the experiment and TDSE calculations for different energies (blue dashed lines in Figs. 2d-h), otherwise, the angular offsets are obviously smaller (red solid lines in Figs. 2d-h). These results explicitly demonstrate that the sub-barrier potential interaction during tunnelling indeed takes time, which can be measurable via the energy-resolved attoclock measurements.

The TCCSFA method is based on the Feynman path integral (FPI), in which the probability amplitude of a quantum wave function can be represented as a coherent superposition of contributions of all possible spatio-temporal paths \cite{Feynman1948RMP, FeynmanBOOK}. Tunnelling is a typical nonclassical process where an electron spends imaginary time within the barrier, $t_{s} = t_{r} + \emph{i}t_{i}$ \cite{Popov2005}. The real part of $t_{s}$ is the ionization time, i.e. the moment when an electron exits the barrier along a certain sub-barrier path, $t_{exit}=t_{r}= \textbf{Re}{(t_{s})}$, and the imaginary part is related to the weight of the electron along this path. Since each path is deterministic, we can trace back exactly the ionization time and the probability of the electron along paths contributing to any part of the final PMD we are interested in \cite{Landsman2014,Hofmann2019,Landsman2015}.

Figure 3a shows the evolution of the electron wave packet (EWP) contributing to the third ATI ring during the tunnelling in the classical forbidden region. For each complex path, the ionization time is constant, while the weight changes during the evolution along the imaginary time. Here, the ionization time $t_{r}$ can be at any time, but the probability might be different. This reflects the probability wave nature of a tunnelling electron, where the behaviour is no longer deterministic and causal. In practice, we can explore the statistical average of the ionization time by evaluating all the involved complex paths: $\langle{t_{r}} \rangle=\frac{\sum_{i}W(i)\cdot t_{r}(i)}{\sum_{i}W(i)}$ \cite{Sokolovski1987,Fertig1990}, where $W(i)$ is the weight of each path $i$.

The green solid line in Fig. 3a shows the statistical averaged ionization time during the imaginary time evolution from $t_{s}$ until $t_{exit}$, i.e., $t_{i} = 0$. It oscillates around the peak of the laser field during the tunnelling. At the tunnelling exit, the statistical averaged ionization time $t_{exit} > 0$. This oscillatory behaviour can be seen more clearly in Fig. 3b where we zoom into the region around $t_{r} = 0$. For comparison, we also show the case where all the potential interactions are neglected (the dashed black line). In this case, its ionization time equals $0$ during the imaginary time evolution without any oscillation, indicating instantaneous tunnelling. Similar rules can be found for other ATI rings.


\begin{figure}[ptb]
 \begin{center}
  \includegraphics[width=0.6\textwidth]{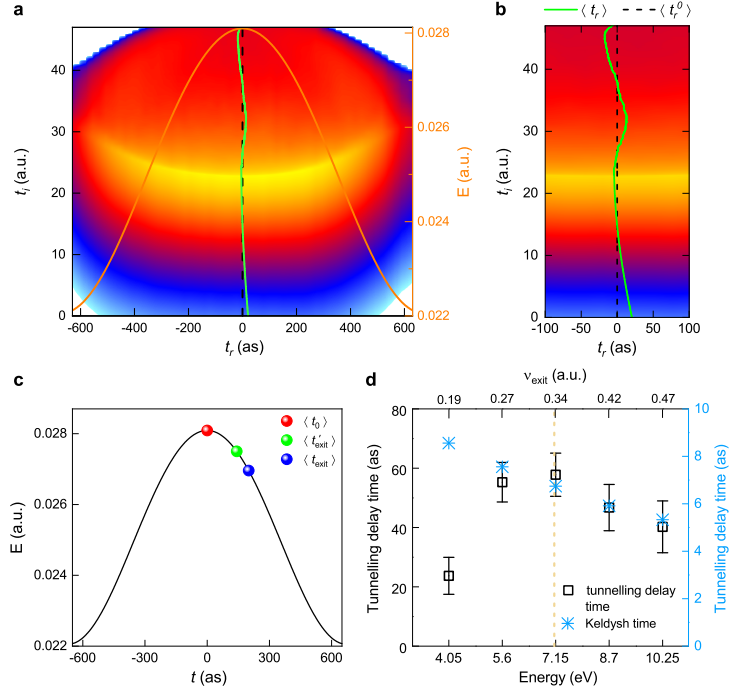}
 \end{center}
 \caption{\emph{\textbf{Fig.3$\mid$ Sub-barrier potential interaction and energy-dependent tunnelling delay time.} \textbf{a,} The probability evolution of EWP along the imaginary time for the third ATI. The horizontal and vertical axes are the real and imaginary parts of the complex time, respectively. The maximum of electric field (solid orange line) is the time $0$. The solid green line presents statistical averaged ionization time $\langle t_{r}\rangle$ at each imaginary time when the potential interactions during and after the tunnelling are included. The dashed black line is the statistical averaged ionization time $\langle{t_{r}^{0}} \rangle$ when the Coulomb interaction is neglected. \textbf{b} is the expansion of \textbf{a} around time $0$. \textbf{c}, The statistical averaged ionization time corresponding to the most probable angular distribution of the third ATI ring when the Coulomb interaction is neglected (red dot), and the Coulomb interaction only after}  }
 \label{fig3}
\end{figure}

\begin{figure}[ptb]
 \caption{\emph{tunnelling is included (green dot), and the Coulomb interaction both during and after the tunnelling are considered (blue dot) in the tunnelling. \textbf{d}, Black squares denote the reconstructed tunnelling delay times from different ATI peaks, and the upper axis shows their corresponding velocities at the tunnelling exit. The light blue stars represent the Keldysh tunnelling time, and the dashed yellow line indicates the velocity matching the effective height of the barrier. The error bars show the standard deviation of statistical errors for reconstructing the tunnelling time from multiple sampling in a range of 8 degrees centred around the most probable angles for each ATI ring.}}
 \label{fig3}
\end{figure}

Then, the following question arises: dose the nonzero statistical averaged ionization time at the tunnelling exit correspond exactly to the tunnelling time, i.e. the time of sub-barrier potential interaction? According to the assumptions in conventional strategies of reconstructing tunneling time\cite{Pazourek2015}, the answer would be 'yes',  because only the tunnelling process could determine the ionization time according to the causality. However, surprisingly, we find that the statistical averaged ionization time of the paths contributing to the highest signal in the PMD is nonzero even when only the Coulomb interaction in the continuum is considered (the green dot in Fig. 3c). This ionization time is slightly smaller than that in TCCSFA simulation but it is obviously larger than that in the case where all the potential interactions are neglected, namely, time 0 (the blue and red dot in Fig. 3c).

In attoclock measurements, the angular offset is measured according to the highest photoelectron signal in the final PMD, which was assumed to be emitted at the peak of the electric field. However, over twenty years ago,  Landauer \emph{et al.} claimed that measuring the tunnelling time by following the peak should 'deserves least attention', because they have noticed that an incident peak of EWP would not turn into a transmitted peak, particularly in the present of strong deformations \cite{1994Landauer, BLtime}. Here, the deformation results from the Coulomb interaction. When Coulomb deformation is completely neglected, the EWP behaves as a plane wave. In this case, an incident peak of the EWP at the tunnelling exit indeed turns into the most probable distribution in the final PMD (black dashed lines in Figs. 3a and 3b, and the red dot in Fig. 3c). However, the outgoing EWP would be divergent owing to the Coulomb interaction in the continuum. Both the final momentum and the phase of each path would be different compared to the case without Coulomb interaction. Owing to the quantum interference, the highest signal in the PMD would arise from a complete different set of paths, where statistical averaged ionization time would be delayed relative to that in the plane wave case (the red dot in Fig. 3c), which means that the Coulomb interaction in the continuum can induce ionization time delay. Obviously, this delay has no correlation with the tunnelling process but is due to the properties of electron wavepacket in the continuum. Hence, only when both wave and particle properties are fully taken into account, the tunnelling dynamics encoded in the attoclock measurements can be fully understood.

The electron tunnelling in the experiment contains potential interactions both within the barrier and in the continuum, and the statistical averaged ionization time of the quantum paths contributing to the most probable signals in PMD is further delayed (the blue dot in Fig. 3c). This additional ionization time delay corresponds to the time of the sub-barrier potential interaction. Close inspection reveals that the impact of the sub-barrier potential interaction on the quantum paths is different with that of Coulomb interaction in the continuum. We find that the sub-barrier potential interaction mainly changes the imaginary part $t_{i}$ of quantum path but keeps the real part $t_{r}$ nearly unchanged, which means that it does not slow down the electron along a certain sub-barrier path, but changes the probability of electron along this path, altering the evolution of the entire tunnelling EWP.

According to the statement of Landuer \emph{et al.} \cite{BLtime,1994Landauer}, following the highest probability signals is the largest drawback inherent in the attoclock measurements. However, this does not mean that attoclock cannot be used to measure the electron tunnelling time. The key point lies in that one should extract the ionization time delay solely induced by the sub-barrier potential interaction. Here, with the wave-particle duality in mind, and based on the fact that the measured energy-resolved most probable angular distribution can
be quantitatively reproduced using the TCCSFA simulation, we propose that the tunnelling delay time can be constructed using the relative ionization time delay between the measurement and the case that neglecting the sub-barrier potential interaction during the tunnelling, i.e. $\tau=\langle t_{exit} \rangle-\langle t_{exit}^{\prime} \rangle$, where $\langle t_{exit} \rangle$ is the corresponding statistical averaged ionization time in the measurement, and $\langle t_{exit}^{\prime} \rangle$ is the statistical averaged ionization time in the case without the sub-barrier potential interaction during the tunnelling (sketched in Fig. 1c).

Figure 3d shows the extracted tunnelling delay time, which ranges from $24$ to $58$ attosecond for ATI rings of different energies.  Interestingly, this energy-dependent trend of the extracted tunnelling delay
time of the electron is consistent with that in the measurements of atom tunnelling \cite{2020nature}. Above the effective barrier height, the trend is in agreement with the improved classical quasi-static model of Keldysh time (see details in the S3), i.e. an electron with higher velocity spends less time during the tunnelling (see the light blue stars in Fig. 3d). Below the effective barrier height, the counterintuitive phenomenon observed in atom tunnelling is also confirmed, i.e. a slower electron with a lower energy has a shorter tunnelling time \cite{2020nature}.


In summary, the tunnelling delay time, i.e., the time of sub-barrier potential interaction, can be explicitly extracted by embracing the wave-particle duality.
Our results not only unify the tunnelling time of atoms and electrons but also gain a deep insight into the tunnelling dynamics of matter waves. Manipulating the sub-barrier potential interaction would provide an unprecedented opportunity to control the tunnelling dynamics in microsystems, which could be used in applications such as manufacturing sub-nanoscale quantum devices at Moore's limit, speeding up enzymatic catalysis, and preventing spontaneous mutations in DNA \cite{Tong2020,Godbeer2015,Ball2011,Lowdin1963,Zhu2020}.


\begin{addendum}
 \item   This work was supported by the National Natural Science Foundation of China (Grant Nos. 91950101, 11627807, 11774130, 11774215, 12074240 and 12004133) and the National Basic Research Program of China (No. 2019YFA0307700), DD and XL acknowledge the support of Science Challenge Project (TZ2018005), and WFY and XHS acknowledge the support of the Department of Education of Guangdong Province (Grant No. 2018KCXTD011) and the Mobility Programme of the Sino-German Center (Grant No. M-0031).

 \item[Author contributions] DD and CW conceived the experiments. XL, CW and YY conducted the measurements, CW, WY, XS, XL and SZ interpreted the results, XL prepared the figures. XL, JL, XS, JC and WY provided the calculations. CW, XL, XS, WY and DD prepared the manuscript. All authors reviewed the manuscript.
 \item[Competing Interests] The authors declare that they have no competing financial interests.
 \item[Correspondence] Correspondence and requests for materials
 should be addressed to Chuncheng Wang (ccwang@jlu.edu.cn), Xiaohong Song~(songxh@stu.edu.cn), Weifeng Yang~(wfyang@stu.edu.cn), or Dajun Ding~(dajund@jlu.edu.cn).
\end{addendum}

\end{document}


\captionsetup{ labelformat=empty}

\maketitle



\noindent\textbf{This PDF file includes:}\\
Section S1. Experimental setup.\\
Section S2. Tunnelling-Coulomb-corrected strong-field approximation (TCCSFA).\\
Section S3. Classical quasi-static tunnelling time.\\
Section S4. Numerical solution of time-dependent Schr\"{o}dinger equation and hydrogen atom.\\
Extended Data Fig.1. Scheme of the experimental setup.\\
Extended Data Fig.2. Energy-resolved attosecond angular streaking and tunnelling delay time of hydrogen atom.\\
References

\newpage
\subsection{Experimental setup.}
Laser pulses were generated by a Spectra-Physics Solstice Ace Ti:Sapphire laser system at a central wavelength of 800 nm. The duration of the laser pulse was 40 fs, and the repetition rate was 1 kHz. Elliptically polarised laser pulses with an ellipticity of 0.79 were produced with the combination of a half-wave plate (HWP) and quarter-wave plate (QWP). The laser pulses were focused into a vacuum chamber by a mirror with a focusing length of 75 mm. A supersonic beam of xenon (seeded in He) was introduced into the chamber and interacted with the focused laser beams. The energy-resolved angular streaking spectra of Xe were measured using Cold-Target Recoil-Ion Momentum Spectroscopy (COLTRIMS) (see Extended Data Fig.1) \cite{setup1,setup2}. The charged particles were accelerated by the combination of a homogeneous electric field (5.3 V/cm) and a magnetic field (7 Gauss), and their time and position could be recorded by multi-channel plates and delay-line detectors. The count rate of photoelectrons was kept lower than 0.2 per pulse to suppress the false coincidence ratio. The peak laser intensity can be calibrated by measuring the ionization yield ratio of Xe$^{2+}$/Xe$ ^{+} $ and the ponderomotive potential shift of the ATI peak, which was (45$\pm$2) TW/cm$^{2}$ in this measurement\cite{Wang2019PRL}.

\subsection{Tunnelling-Coulomb-corrected strong-field approximation (TCCSFA).}
In strong field ionization, the transition matrix element is given by
\begin{equation}
M\left(\mathbf{p}\right)=-i \int_{t_{0}}^{\infty} d \tau\left\langle\Psi_{f}|\mathbf{r} \cdot \mathbf{E}(\tau)| \psi_{0}\right\rangle e^{-i S(\mathbf{p},\tau)},
\end{equation}
the action $S$ is a multi-dimensional complex-valued function which is comprised of three components
\begin{equation}
S\left(\mathbf{p}, t\right)=S_{\mathrm{V}}\left(\mathbf{p},t\right)+S_{\mathrm{C}}\left(\mathbf{p}, t\right)-I_{p} t.
\end{equation}
Here the first component is the action accumulated by the electron in the laser field only
\begin{equation}
S_{\mathrm{V}}\left(\mathbf{p}, t\right)=\frac{1}{2} \int_{t}^{\infty} d \tau[\mathbf{p}+\mathbf{A}(\tau)]^{2},
\end{equation}
and the second one is the action accumulated due to the electron with the cores
\begin{equation}
S_{\mathrm{C}}\left(\mathbf{p}, t\right)=\frac{1}{2} \int_{t}^{\infty} d \tau V[\mathbf{r}(\mathbf{p},\tau)],
\end{equation}
where $V[\mathbf{r}(\mathbf{p},\tau)]$ is the potential of atom evaluated along the electron's laser-driven quantum trajectory
\begin{equation}
\mathbf{r}(\mathbf{p}, \tau)=\int_{t}^{\tau} d \tau^{\prime}\left(\mathbf{p}+\mathbf{A}\left(\tau^{\prime}\right)\right).
\end{equation}
The saddle point equation follows from Eq.(2) that
\begin{equation}
\left.\frac{\partial S(\mathbf{p}, t)}{\partial t}\right|_{t=t_{s}}=\left.\frac{\partial S_{\mathrm{V}}(\mathbf{p}, t)}{\partial t}\right|_{t=t_{s}}+\left.\frac{\partial S_{\mathrm{C}}(\mathbf{p}, t)}{\partial t}\right|_{t=t_{s}}-I_{p}=0.
\end{equation}
Here the solution of the saddle point equation is a complex time $t_{s}=t_{r}+i \cdot t_{i}$. In the tunnelling picture, the real part $t_{r}$ of the complex time denotes the ionization time when an electron reaches the tunnelling exit and enters the continuum. The imaginary part $t_{i}$ is closely related to the weight of electron along a certain path. The difficulty of solving the  accurate saddle point equation (6) lies in that the potential interaction term $S_{\mathrm{C}}$ is not an analytical function of the time $t$. The term $S_{\mathrm{C}}$ is usually neglected in the strong field approximation (SFA) theory \cite{Keldysh1965,Faisal1973,Reiss1980,Lewenstein1994}. Thus the reduced saddle point equation is
\begin{equation}
\left.\frac{\partial S_{\mathrm{V}}(\mathbf{p}, t)}{\partial t}\right|_{t=t_{s}}-I_{p}=0,
\end{equation}
which can be easily solved. However, as we have mentioned, the potential interaction should be in the action $S$ which is a multi-dimensional complex-valued function. It is a formidable challenge to solve Eq.(6). In our work, we provide a self-consistent method to solve the saddle point equation with the potential interaction.

Because the integrand of $S_{\mathrm{V}}$  is an analytic function of complex time $t$, see Eq.(3), we can get
\begin{equation}
\left.\frac{\partial S_{\mathrm{V}}(\mathbf{p}, t)}{\partial t}\right|_{t=t_{s}}=-\frac{1}{2}[\mathbf{p}+\mathbf{A}(\tau)]^{2}.
\end{equation}
However, $V[\mathbf{r}(\mathbf{p},\tau)]=Z/\mathbf{r}(\mathbf{p},\tau)$, with $Z$ the effective charge of the ionic core, is dependent on integral-path because it's not an analytic function in the multi-dimensional complex plane when $\mathbf{r}$ is close to zero. Thus, the differential of Eq.(4) cannot be directly deduced. To solve this problem, we divide the integral of Eq.(4) into two parts in complex-time plane
\begin{equation}
S_{\mathrm{C}}(\mathbf{p}, t)=\int_{t}^{\infty} d \tau V[\mathbf{r}(\mathbf{p}, \tau)]=\int_{{\rm I}_{1}} d \tau V[\mathbf{r}(\mathbf{p}, \tau)]+\int_{{\rm I}_{2}} d \tau V[\mathbf{r}(\mathbf{p}, \tau)],
\end{equation}
where the integral-path ${\rm I}_{1}$ is along the imaginary time with a constant $t_{r}$, while $t_{i}$ is changing until $t_{i}=0$. The integral-path ${\rm I}_{2}$ is along real time with $t_{i}=0$. Then the potential interaction term in the saddle point equation would be
\begin{equation}
\left.\frac{\partial S_{\mathrm{C}}(\mathbf{p}, t)}{\partial t}\right|_{t=t_{s}}=\left.\left[\frac{\partial}{\partial t} \int_{{\rm I}_{1}} d \tau V[\mathbf{r}(\mathbf{p}, \tau)]+\frac{\partial}{\partial t} \int_{{\rm I}_{2}} d \tau V[\mathbf{r}(\mathbf{p}, \tau)]\right]\right|_{t=t_{s}},
\end{equation}
where the first term can be written as
\begin{equation}
\begin{aligned}
\left.\frac{\partial}{\partial t} \int_{{\rm I}_{1}} d \tau V[\mathbf{r}(\mathbf{p}, \tau)]\right|_{t=t_{s}} &=\lim _{\Delta \tau \rightarrow 0} \frac{\int_{t_{i}+\Delta \tau}^{0} V\left[\mathbf{r}\left(\mathbf{p}, t_{r}+i \cdot \tau\right)\right] d \tau-\int_{t_{i}}^{0} V\left[\mathbf{r}\left(\mathbf{p}, t_{r}+i \cdot \tau\right)\right] d \tau}{\Delta \tau} \\
&=-V\left[\mathbf{r}\left(\mathbf{p}, t_{s}\right)\right],
\end{aligned}
\end{equation}
and the second term can be written as
\begin{equation}
\begin{aligned}
\left.\frac{\partial}{\partial t} \int_{{\rm I}_{2}} d \tau V[\mathbf{r}(\mathbf{p}, \tau)]\right|_{t=t_{s}} &=\lim _{\Delta \tau \rightarrow 0} \frac{\int_{t_{r}+\Delta \tau}^{\infty} V[\mathbf{r}(\mathbf{p}, \tau)] d \tau-\int_{t_{r}}^{\infty} V[\mathbf{r}(\mathbf{p}, \tau)] d \tau}{\Delta \tau} \\
&=-V\left[\mathbf{r}\left(\mathbf{p}, t_{r}\right)\right].
\end{aligned}
\end{equation}
Substituting Eqs.(8)-(12) to Eq.(6), then we can get the saddle point equation with the potential interaction
\begin{equation}
\frac{1}{2}\left(\mathbf{p}+\mathbf{A}\left(t_{s}\right)\right)^{2}
+V\left[\mathbf{r}\left(t_{s}\right)\right]
=-I_{p}-V\left[\mathbf{r}\left(t_{r}\right)\right].
\end{equation}
Here, $\mathbf{r}$ is the position of the electron, $V[\mathbf{r}(\mathbf({t_s}))]=-Z_{eff}/|\mathbf{r}({t_s})|$ is the potential energy at the tunnelling entrance and $V[\mathbf{r}(\mathbf({t_r}))]=-Z_{eff}/|\mathbf{r}({t_r})|$ is the potential energy at the tunnelling exit, where $Z_{eff}=\sqrt{2 I_{p}}$ is the effective charge of the ionic core.

Physically, the saddle point equation describes energy conservation in tunnelling with regard to SFA, in which the energy was assumed to change from the ground state with energy $-I_{p}$ before tunnelling to the continuum state with zero energy after tunnelling (see Fig. 1a in the main text). Hence, when $I_{p}$ was employed to describe the energy difference of the tunnelling ionization in conventional semiclassical methods, such as SFA and Coulomb-corrected strong-field approximation (CCSFA), it had been assumed that the electron was totally free and could not experience the effect of the ionic potential after tunnelling. However, when emerging at the exit after penetrating a finite potential barrier, the EWP is not totally free but still experiences an attractive force owing to the potential at the tunnelling exit with $V\left[\mathbf{r}\left(t_{r}\right)\right]$ (see Fig. 1b in the main text). As a result, the variation of the potential energy before and after tunnelling should be equal to $V\left[\mathbf{r}\left(t_{r}\right)\right]+I_{p}$. According to the energy conservation, we obtain the saddle point equation with the potential interaction, i.e., Eq.(13).

After the tunnelling, the subsequent motion of the electron in the continuum is governed by the combined action of the laser and Coulomb fields. The initial distribution for an electron emerging at the tunnelling exit in the continuum, including its initial velocity and position, and the weight of each electron trajectory can be obtained from the saddle point equation with potential interaction. Then, the motion of the electron in the real-time propagation is determined by Newton's equations:
\begin{equation}\left\{\begin{array}{l}
\dot{\mathbf{p}}(t)=-\nabla_{\mathbf{r}} V[\mathbf{r}(t)]-\mathbf{E}(t) \\
\\
\dot{\mathbf{r}}(t)=\mathbf{p}(t)
\end{array}.\right.
\end{equation}

In this study, we employed a fourth-order Runge-Kutta method to solve Eq. (6). The phase in real time is
\begin{equation}
S\left(\mathbf{p},t_{r}\right)=\int_{t_{r}}^{\infty}\left(\frac{1}{2} \mathbf{v}^{2}(\tau)-\frac{Z_{eff}}{|\mathbf{r}(\tau)|}+I_{p}\right) d\tau,
\end{equation}
where $\mathbf{v}(\tau)$ is the instantaneous velocity of the electron. As the Coulomb interactions both during the tunnelling and in the continuum have been included, we call this method TCCSFA.

\subsection{Classical quasi-static tunnelling time.}

The work performed by the conservative Coulomb force during the tunnelling is
\begin{equation}
\int_{{t_s}}^{{t_r}}{F_C(r)} dr=V[\mathbf{r}({t_r})]-V[\mathbf{r}({t_s})],
\end{equation}
where $F_C(r)$ is the Coulomb force, and $V[\mathbf{r}({t_r})]=-Z_{eff}/|\mathbf{r}({t_r})|$ and $V[\mathbf{r}({t_s})]=-I_{p}$ are the potential energy at the tunnelling exit and at the entrance of the barrier, respectively. Owning to ${r}({t_s})\approx0$, we obtain $\widetilde{F_C}\cdot |\mathbf{r}({t_r})|\approx -Z_{eff}/|\mathbf{r}({t_r})|+I_{p}$. Therefore, the averaged Coulomb force during the tunnelling is
\begin{equation}
\widetilde{F_C}\approx-Z_{eff}/|\mathbf{r}({t_r})|^2+I_{p}/|\mathbf{r}({t_r})|.
\end{equation}

Considering that tunnelling is a dynamic process with a very short time scale, we can assume that the laser field has not drastically changed and appraise the tunnelling time with an improved classical quasi-static model \cite{Ivanov2005}. The velocity of the electron travelling through the barrier created by the averaged electric and Coulomb fields changes as

\begin{equation}
v(t)=v_{\mathrm{en}}+({F_L}+\widetilde{F_C})\cdot t.
\end{equation}
Here, $v_{\mathrm{en}}$ is the velocity at the entrance under the barrier within the classical model and $F_L$ is the peak laser field. Substituting the velocity at the entrance and at the exit in the above equation, we can obtain the classical tunnelling time:
\begin{equation}
\tau_{\mathrm{tunnelling}}\approx\frac{v({t_r})-v_{\mathrm{en}}}{{F_L}-Z_{eff}/|\mathbf{r}({t_r})|^2+I_{p}/|\mathbf{r}({t_r})|}.
\end{equation}
Therefore, substituting the statistical averaged velocity $\langle{v({t_r})}\rangle$ and positions $\langle{\mathbf{r}(t_r)}\rangle$ at the tunnelling exit from the TCCSFA simulation and $v_{\mathrm{en}}=\sqrt{2 I_{p}}$ into the above equation, we qualitatively determine the classical tunnelling time $\tau_{\mathrm{tunnelling}}$.

The effective height of the potential barrier is
\begin{equation}
\Gamma=\frac{\int_{{r_{\mathrm{en}}}}^{{r_{\mathrm{exit}}}}{(-Z_{eff}/|\mathbf{r}({t})|+\mathbf{F_L}\cdot \mathbf{r}({t})+I_{p})} dr}{{r_{\mathrm{exit}}}-{r_{\mathrm{en}}}},
\end{equation}
where $r_{\mathrm{en}}$ and $r_{\mathrm{exit}}$ are the position at the entrance and at the exit of the barrier, respectively.

\subsection{Numerical solution of time-dependent Schr\"{o}dinger equation and hydrogen atom}

The time-dependent Schr\"{o}dinger equation (TDSE) is numerically solved by the Qprop package combined with the time-dependent surface flux (t-SURFF) \cite{Mosert2016,Tulsky2020,Tao2012}. A model atomic potential of Xenon is employed:
\begin{equation}
V(r)=-\frac{1+a_1 e^{-a_2 r}+a_3 e^{-a_4 r}+a_5 e^{-a_6 r}}{r},
\end{equation}
where the parameters are given from ref. \cite{Milosevic2010}. The numerical solutions of the TDSE, which have been reproduced successfully in several experiments \cite{Tulsky2020,Song2018}, together with the measurements in this work, serve as the benchmarks for checking the validity of the semiclassical CCSFA and TCCSFA simulations.

The hydrogen atom is a single-electron system where the multielectron effect was ruled out, which allows an exact numerical solution of the TDSE. We then compare the numerical results of the TDSE with the TCCSFA calculations, and again find very good agreement for ATI rings of different energies (see Extended Data Fig. 2a-e). The reconstructed tunnelling delay time of hydrogen (see Extended Data Fig. 2f) shows energy-dependent behavior similar to that of xenon, which confirms that the delay time is induced by the tunnelling of the electron rather than the multielectron dynamics (see main text).


\begin{figure}
	\centering
	\includegraphics[width=0.8\textwidth]{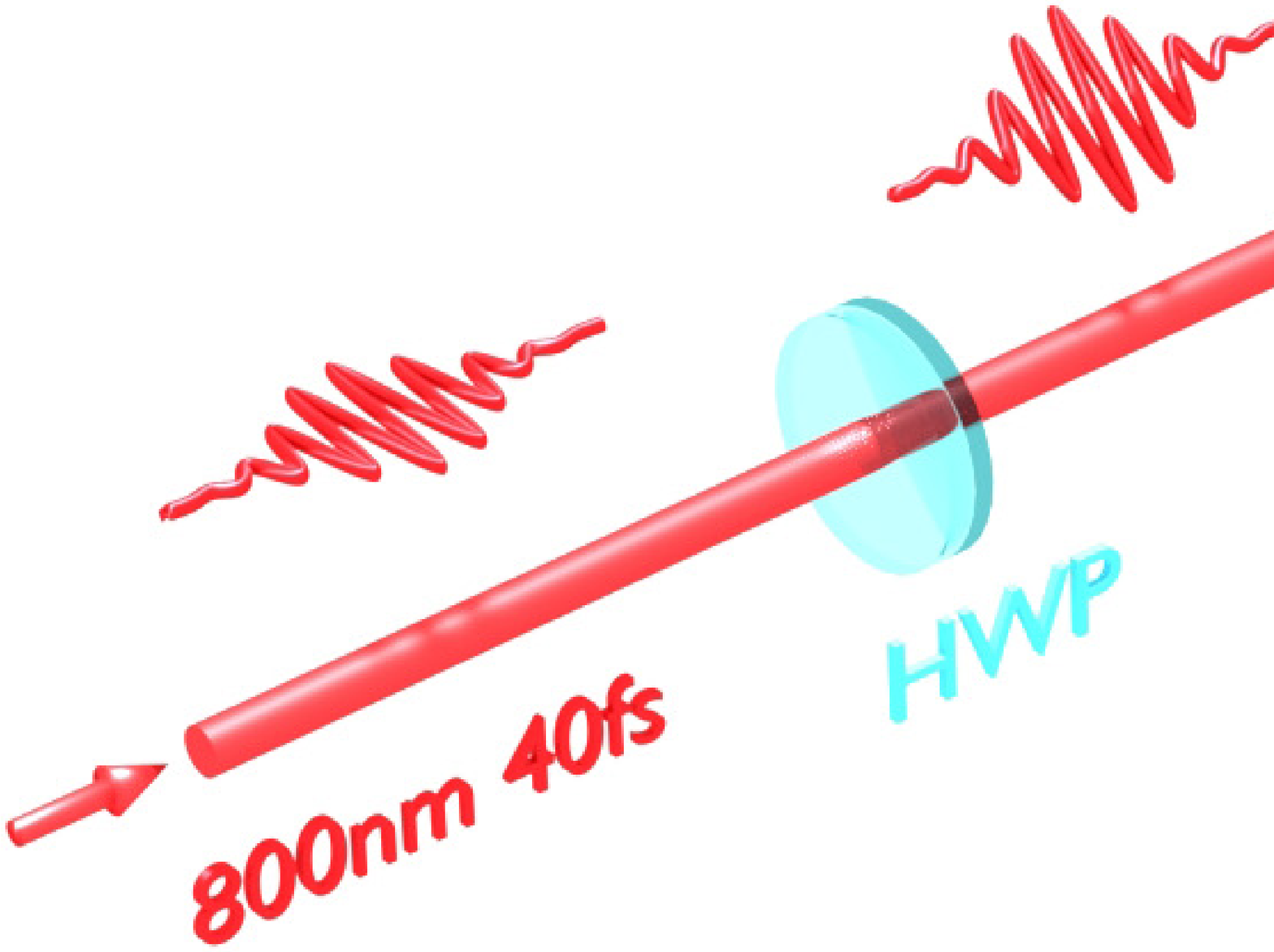}
	\caption{\textbf{Extended Data Fig.1$\mid$Scheme of the experimental setup.} The energy-resolved ATI spectra of Xe irradiated by elliptically polarised laser fields are measured in coincidence with the parent ions using COLTRIMS. A combination of a half wave plate (HWP) and quarter wave plate (QWP) is used to produce the elliptical polarised laser pulses. The produced ions and electrons are accelerated and guided by the electric (E) and magnetic(B) fields and then detected by two oppositely positioned sensitive delay-line detectors.}
	\label{figure4}
\end{figure}

\begin{figure}
	\centering
	\includegraphics[width=\linewidth]{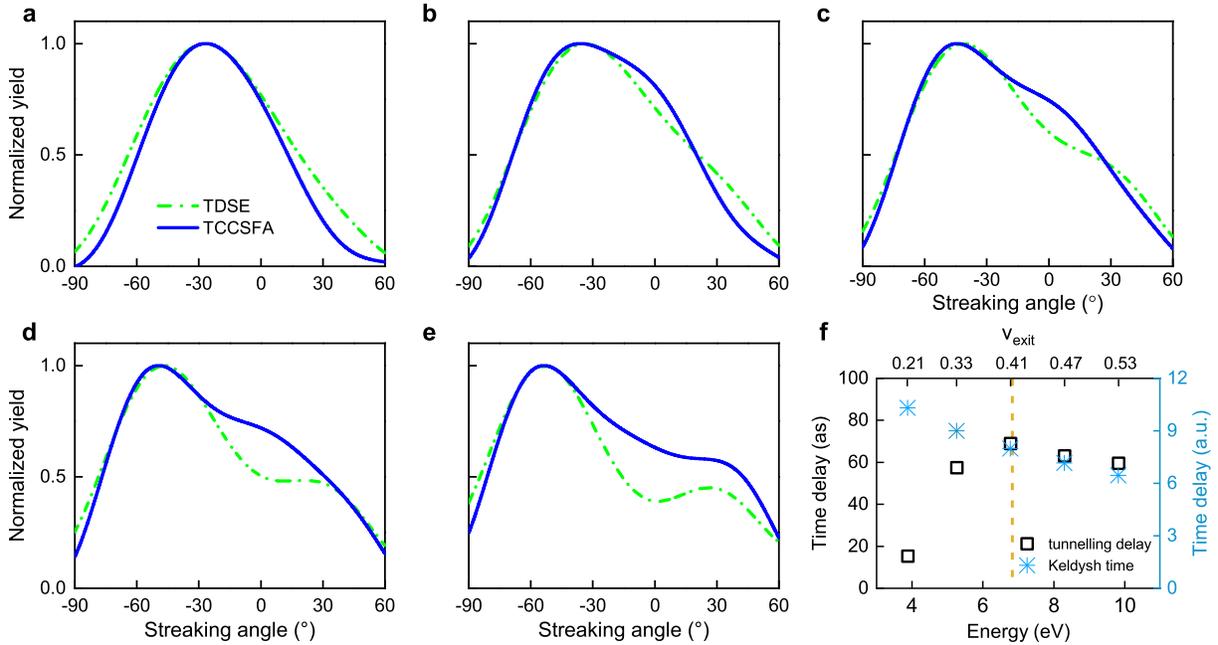}
	\caption{\textbf{Extended Data Fig.2$\mid$Energy-resolved attosecond angular streaking and tunnelling delay time of hydrogen atom.} \textbf{a-e}, The simulated and calculated angular distributions for ATI rings with different energies acquired from the numerical solution of 3D-TDSE (green dash-dotted line) and the TCCSFA simulation (blue solid line), whose peak angles are in good agreement. \textbf{f}, The reconstructed tunnelling delay time for different energies of ATI. The blue stars are the Keldysh tunneling time and the dashed yellow line indicates the velocity matching the effective height of the barrier (see the Methods).}
	\label{figure5}
\end{figure}
